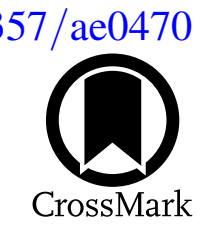

# Analysis of Short-term Solar Activity Variability and Estimating the Timings of the Next Enhanced Bursts

Juie Shetye[1] and Mausumi Dikpati[2]
[1] Department of Astronomy, New Mexico State University, Las Cruces, NM 88003, USA; jshetye@nmsu.edu
[2] High Altitude Observatory, NSF-NCAR, 3080 Center Green Drive, Boulder, CO 80301, USA

## Abstract

We present a novel hybrid forecasting strategy combining numerical, statistical, and machine learning–based forecasting to detect the occurrence of the next enhanced solar activity bursts. These enhanced bursts are called "space weather seasons," which occur on intermediate timescales (6–18 months). Monthly smoothed sunspot number (SSN) data from 1878 to 2025 are analyzed using Gaussian fitting techniques to identify burst events and their properties such as amplitude and duration. The SSN data are divided into training, test, and forecast, which shows hindcast and forecast. Each hemisphere is modeled via a seasonal autoregressive integrated moving average approach, refined with an asymmetric Gaussian override to capture rapid burst rise and gradual decay, and burst amplitudes and duration are predicted using a random forest regression model. This hybrid approach successfully hindcasts burst timing in between 2024 November and 2025 May, with a peak SSN of ∼70 around 2025 March for the Northern Hemisphere. The next burst in the Northern Hemisphere is forecast to be in 2025 December with a slightly lower SSN of 60. By contrast, the Southern Hemisphere shows relatively complicated behavior, where the bursts show multiple amplitudes starting approximately in 2024 October and ending in 2025 October. The main burst shows an amplitude of 130 SSN. The next burst in the Southern Hemisphere is forecast to occur approximately in 2025 December. Combining SSN properties in both hemispheres, we find that the total SSN is mainly influenced by a stronger cycle in the Southern Hemisphere.

## 1. Introduction

Cyclic solar activity variability and bursts, occurring on multispatiotemporal scales, have a profound influence on our technological society. Decadal timescale solar activity, called the solar cycle (SC), occurs with average 11 yr periods. Although an SC progresses roughly sinusoidally, it does not follow a nice smooth sinusoid pattern. Instead, each SC progresses in the form of a quasi-annual-type enhanced activity burst followed by a relatively quiet period (see, e.g., Figure 2(a) of M. Dikpati et al. 2017). These short-term, quasi-annual variabilities in solar activity are widely known as Rieger-type periodicity (E. Rieger et al. 1984) and quasi-biennial oscillations (A. Vecchio & V. Carbone 2009). Many subsequent analyses confirm these periods in solar activity (E. Gurgenashvili et al. 2016) including hemispheric asymmetry (E. Gurgenashvili et al. 2017) and in many other indices, such as solar irradiances (E. Gurgenashvili et al. 2022).

In this study, a "burst" refers to a quasiperiodic enhancement in solar activity, characterized by a rapid increase in the smoothed sunspot number (SSN) followed by a gradual decline, typically lasting 6–18 months. These quasiperiodic short-term solar activity bursts on the timescales of 6–18 months constitute the so-called space-weather "seasons" (S. W. McIntosh et al. 2015, 2017). They are called the "seasons" of space weather primarily because the biggest class flares and coronal mass ejections (CMEs) have been observed to occur during the time span of enhanced bursts of activity (see, e.g., Figure 2 of M. Dikpati & S. W. McIntosh 2020). A. Lobzin et al. (2012) reported the existence of such periods in type III radio bursts. "Seasonal" (quasi-annual) enhanced activity bursts and associated energetic events that cause space weather occur in all SCs, irrespective of the decadal cycle strength. For example, during the medium-strength Cycle 23, Halloween storms with an X45 class flare occurred just after the peak (M. Dikpati et al. 2021), and also in the low-strength Cycle 24, 2017 September, solar storms occurred in the late declining phase, producing X13.8 flares (B. Raphaldini et al. 2023), both during the enhanced activity bursts. The current Cycle 25, which is in its peak, created several major solar storms including a significant one, i.e., the Mother's Day or Gannon superstorm in 2024 May (M. Dikpati et al. 2025; K. S. Paul et al. 2025). Thus we see that the amplitude and duration of these enhanced activity bursts are equally as important as the decadal solar activity cycle's strength and duration, and perhaps more.

What causes these short-term quasiperiodic solar activity variability and quasi-annual "seasons" of space weather? There have been various attempts to understand and physically interpret these quasi-annual activity bursts. One class of models is the magnetohydrodynamics (MHD) shallow-water tachocline models, which have been employed during the past two decades (P. A. Gilman 2000; T. V. Zaqarashvili et al. 2007). In these models, short-term, Rieger-type solar activity variability is shown to occur due to global MHD instabilities, as a result of which MHD Rossby waves get generated. The spot-producing magnetic fields of the Sun most likely get generated in the subadiabatically stratified tachocline. T. V. Zaqarashvili et al. (2010) have shown that the unstable harmonics of fast MHD Rossby waves for a 10 kG magnetic field have periods of about 5 months, whereas the periods increase (decrease) with field strengths for symmetric (antisymmetric) modes.

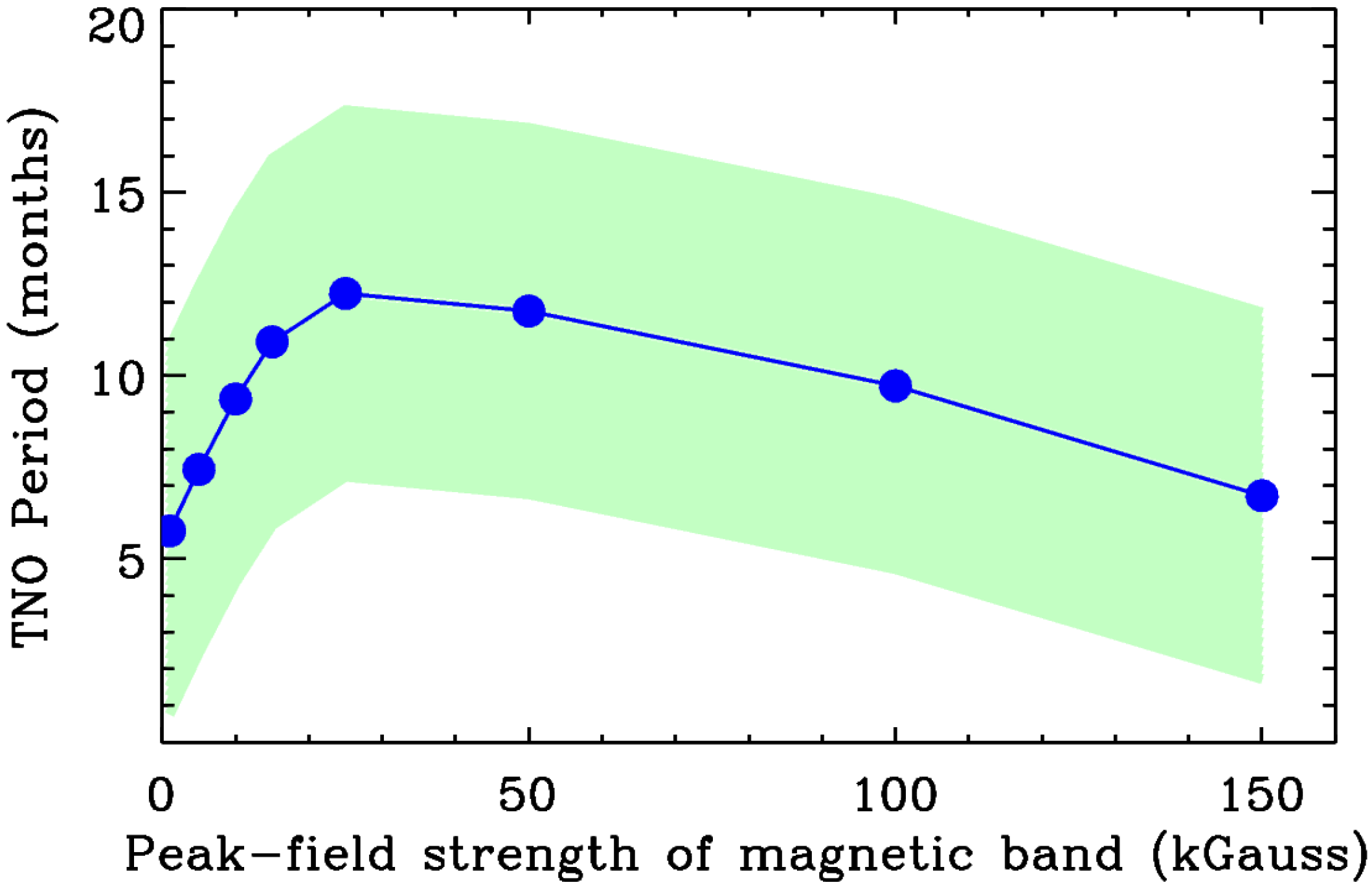

**Figure 1.** TNO period as function of field strength is displayed as the blue curve. In addition to field strength, a wide range of parameters like tachocline differential rotation, the amount of possible subadiabiticity, and latitude location of the magnetic band give rise to changes (increase or decrease) in the TNO period; the range of TNO periods has been captured by the spread displayed in the light green shade. The TNO periods follow the $x$-axis going from left to right during the rising phase since the sunspot number (as well as spot area and magnetic flux), which are a proxy of tachocline magnetic fields, increase sinusoidally.

In the nonlinear interactions among Rossby waves, magnetic fields, and differential rotation in an MHD shallow-water model, the periods of short-term solar variability come from dynamical exchanges among the reference state and perturbation kinetic, magnetic, and potential energies. These are called tachocline nonlinear oscillations (TNOs; M. Dikpati et al. 2017). For a wide range of parameters, namely the tachocline shell thickness, the subadiabaticity therein, the field strength of toroidal magnetic bands, and their latitude locations, the TNO periods were found to be within the range of 2 and 20 months (M. Dikpati et al. 2018b). The TNOs could explain the short-term quasiperiodic solar activity variability in that range too. Figure 1 shows the plausible range of the TNO periods from an MHD shallow-water tachocline model, expected to be observed in surface solar activity variability.

While in the solar context, the shallow-water models have been applied to the thin tachocline layer, the application of this class of models has also been studied in the thin surface layer, namely the supergranulation layer (M. Dikpati et al. 2022, 2024), and quasiperiodic nonlinear exchanges of energies among various energy reservoirs occur there too. Thus, if the origin of these quasi-annual periods in solar variability is near the surface, instead of a deep layer, this physical foundation is a valid mechanism there too. Another class of models that have been used to physically explain the causes of short-term variability in an SC are dynamo models. It has been shown by F. Inceoglu et al. (2019) that the interplay between flows and magnetic fields in terms of back reaction can produce short-term variability in solar activity.

These MHD and dynamo models, though they can provide a foundation for the physics behind the short-term quasi-annual and quasi-biennial periods in solar activity variability, cannot predict when the next activity burst would occur, with what amplitude, and whether the bursts would occur in both hemispheres or in one first and then in the other. This is because the exact parameter estimation for the models mentioned above is extremely difficult. Parameter estimation is possible via data assimilation (see, e.g., S. Dash et al. 2024) and/or machine learning techniques, which use big data to couple data and model. Such an approach is under progress and will be employed in the future.

Currently what is the best was to predict upcoming bursts with a few months' lead time? Can we actually infer from the analysis of data trends and/or models when the next enhanced activity burst will come? This certainly would give enough lead time, of a few weeks to months, to prepare for hazardous space weather impacts on the terrestrial system. So far flares/CMEs are predicted only a few hours before they occur. C. J. Schrijver et al. (2015) pointed out in the COSPAR road map that a lot of effort has been given to understanding and forecasting solar activity on very short timescales (hours to days), like flares/CMEs, and also the decadal SC amplitude and timing (11 yr), but solar activity occurring on intermediate scales (weeks to months) has not received much effort. Recently solar activity bursts occurring on a 6–12 month timescale have been identified to be one of the important goals for making further progress in space-weather events and their impact (see, e.g., page 529 of the decadal survey material).[3] Motivated by this need, our study aims to identify when these intermediate-scale bursts will occur, providing weeks to

[3] https://nap.nationalacademies.org/read/27938/chapter/14#528

months of advanced notice about upcoming "space-weather seasons."

In this manuscript, we develop a new method that combines mathematical formulation and statistical analysis with machine learning techniques to forecast enhanced solar activity bursts occurring every 2–24 months. We obtain monthly sunspot data collected from 1878 to 2025 from the World Data Center (WDC) Sunspot Index and Long-term Solar Observations (SILSO) at the Royal Observatory of Belgium, Brussels (F. Clette et al. 2014). We analyze the data using Gaussian smoothing methods to isolate these short-term bursts from the longer-term SC trend, and predict when the next solar activity bursts will occur. We also note that the main effects of the SC have dominated in the Southern Hemisphere during the recent period.

The paper is structured in the following way. Step 1 of the burst detection algorithm based on Gaussian fits is described in Section 2. Step 2 involves the statistical prediction methodology, employing the seasonal autoregressive integrated moving average (SARIMA) for forecasting bursts is described in Section 3. The final step (Step 3) involves the application of the random forest (RF) machine learning model is described in Section 4. The results of this hybrid approach are described in Section 5, followed by the concluding remarks in Section 6.

## 2. Detection of Enhanced Solar Activity Bursts

In this study, we utilize hemispheric SSN data obtained from the WDC-SILSO at the Royal Observatory of Belgium in Brussels (F. Clette et al. 2014). We use daily sunspot data covering the period from 1878 to 2025 and calculated the average monthly values to represent the SSNs.

We conduct burst identification and characterization using a Gaussian smoothing function implemented with `gaussian_filter1d` in `scipy`.[4] We apply a short-term Gaussian kernel with a FWHM of 4 months to the monthly hemispheric SSN to emphasize subannual structure while filtering month-scale variability. We define an enhanced burst as a short-duration (∼6–18 months), single-lobe enhancement that stands out on the 4 month smoothed series. Candidate bursts are first flagged by visual inspection of the 4 month curve and retained when they satisfy three reproducible criteria: (i) SSN peak $\geqslant 10$ SSN, (ii) minimum separation between neighboring peaks $\geqslant 2$ months, and (iii) an apparent width within 6–18 months (estimated from the FWHM). Each selected interval is then quantified by fitting a Gaussian profile (see Equation (1)) to the 4 month curve in a local window around the peak, yielding the peak time, fitted amplitude, and FWHM; the burst start and end are reported at the half-maximum points. The SILSO cycle minimum yields no accepted bursts in either hemisphere, because minimum-phase oscillations do not simultaneously exceed $A_k \geqslant 10$ and $6 \leqslant W_k \leqslant 18$ months. See the minima regions in Figure 2.

To justify the 4 month window, we compare 2, 3, 4, 6, and 8 month Gaussian smoothers across SC12–25 (Appendix A). Shorter windows (2–3 months) oversegment month-scale undulations, whereas longer windows (6–8 months) broaden or merge nearby features; the 4 month smoother provides the most stable peak timings and high peak prominence while preserving the ∼6–18 month morphology used for selection. See example comparisons for 3–6 month windows plotted in Figure 2.

Next, we select two points on the plot to define the beginning and end of a potential burst region. This interactive selection has been implemented using the `ginput` (see footnote 4) function. Within this user-defined area, we fit a Gaussian function (see Equation (1) below). This modeling approach is chosen because observed bursts frequently exhibit a Gaussian shape, and the linear background accounts for any underlying trend in the SSN during the burst. The model function, $f(x)$, is defined as follows:

$$f(x) = a \cdot \exp\left(-\frac{(x-b)^2}{c^2}\right) + d \cdot x + e, \tag{1}$$

where $x$ represents time in decimal years, $a$ is the amplitude of the Gaussian component (burst height above the background), $b$ is the center (peak position) of the Gaussian (time of burst maximum), $c$ is the standard deviation of the Gaussian (related to burst width), $d$ is the slope of the linear background, and $e$ is the intercept of the linear background.

The fitting process utilizes nonlinear least squares to determine the optimal parameters ($a$, $b$, $c$, $d$, and $e$) that minimize the difference between the model function (Equation (1)) and the observed SSN data. Initial parameter estimates are derived from the user-selected region. The initial amplitude ($a$) is estimated as the difference between the maximum SSN within the region and the average of the endpoints. The center ($b$) value is estimated using the $x$-value at the midpoint of the interval and width ($c$) from the endpoints. Similarly, the slope ($d$) is determined as the slope of the line. Finally, the intercept ($e$) is derived from the region. Note that a Gaussian fit returns the same values for $a$, $b$, $c$, $d$, and $e$ irrespective of the exact start point and endpoint selected by the user within an enhanced burst pattern.

Once the Gaussian fit is obtained, several key properties are extracted to characterize the burst. The peak position is given by the parameter $b$, the time of the fitted Gaussian's maximum. The peak SSN is the value of $(f(x))$ at $x = b$. The amplitude is taken as the maximum SSN value within the user-selected interval. $\mathrm{FWHM} = 2\sqrt{2\ln(2)}\,|c|$, which is converted to days using a factor of 365 days yr$^{-1}$. The duration is the difference between the end and start dates (in days). A plot displaying the original data, selected region, and the fitted Gaussian curve is shown to visually assess the quality of the fit.

Examples of the Gaussian fitting procedure for bursts in both hemispheres are shown in Figure 3. The interactive nature of the burst identification method allows flexibility in accommodating variations in burst shape and background trends, enabling precise extraction of key burst parameters such as the start date, end date, peak position, amplitude, FWHM, and duration. These parameters provide a quantitative basis for subsequent statistical analysis and forecasting model development as detailed in Section 3. Using this method, 144 bursts are identified in the Northern Hemisphere. The peak SSN ranges between 10 and 191 and the mean value is 67. Bursts last a median of 256 days, spanning from 11 to 547 days, and their median FWHM is 135 days. These figures indicate frequent moderately intense events with compact temporal envelopes. One hundred and thirty-seven bursts are identified in the Southern Hemisphere. The peak amplitude varies from 10 to 158 SSN, with a mean near 69. Burst lifetimes center around ≈228 days, bounded by 26 and 646 days, and the median FWHM is 123 days. The Southern Hemisphere, therefore, displays approximately the same event count as the Northern Hemisphere, with a slightly higher typical amplitude and a comparable duration.

[4] https://docs.scipy.org/doc/scipy/reference/generated/scipy.ndimage.gaussian_filter1d.html

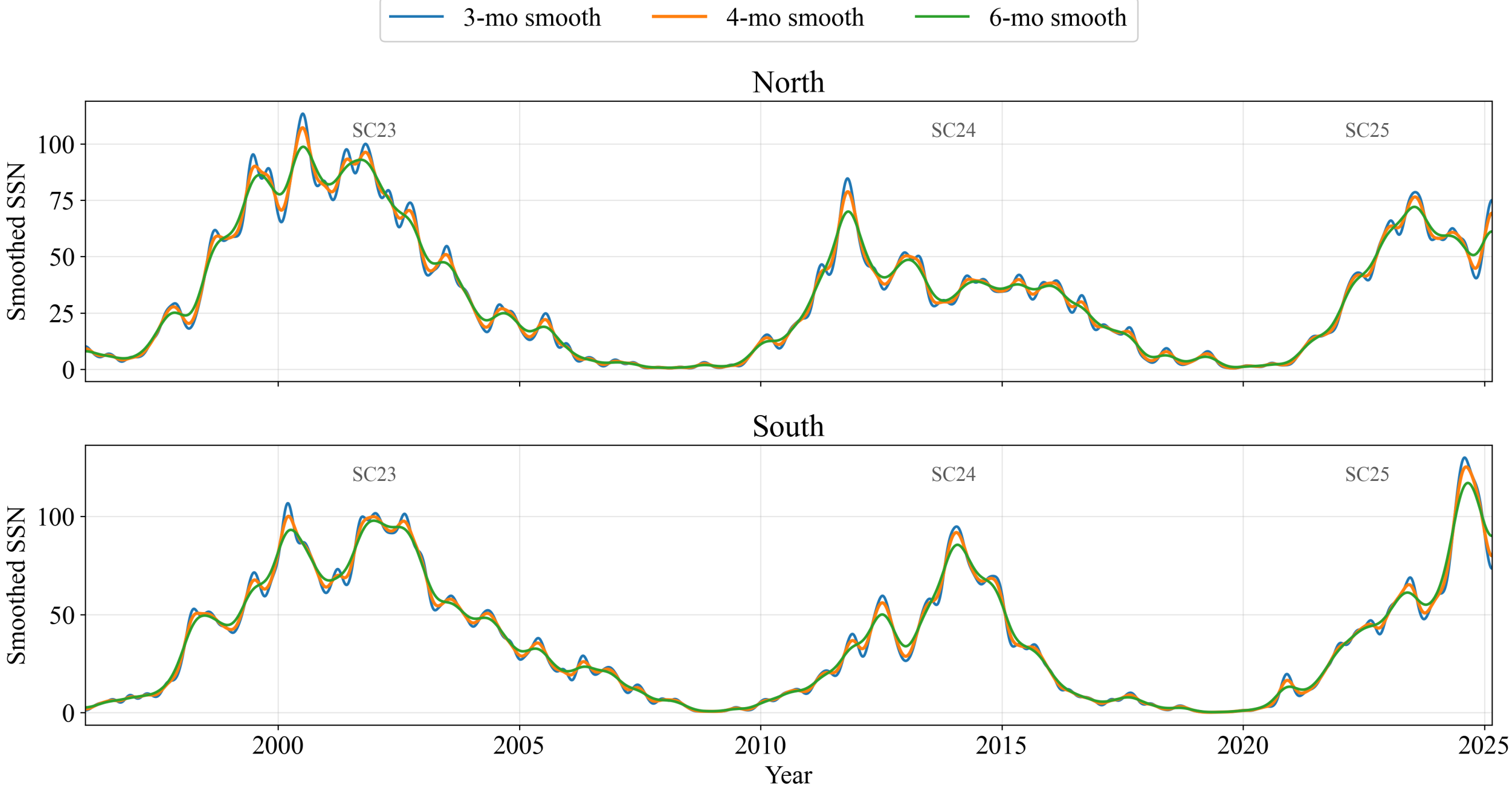

**Figure 2.** Hemispheric SSN from the start of SC23 to 2025 for the north (top) and south (bottom). Curves show Gaussian smooths with FWHMs of 3, 4, and 6 months; no detrending is applied. Shaded bands indicate SC23–SC25.

### 2.1. Results of the Gaussian Analysis

To understand the behavior of solar bursts, we use kernel density estimation (KDE) on the complete 1878–2025 time interval, as a nonparametric method to visualize the distributions of burst amplitude and duration without assuming any specific parametric form. KDE provides a smooth and continuous representation of empirical data, effectively highlighting important distributional features such as peaks, skewness, and tails. This allows for intuitive comparisons across different hemispheres and phases of the SC.

Mathematically, KDE represents each observed burst measurement ($x_i$) by placing a Gaussian kernel centered on that data point, given by

$$K_h(x - x_i) = \frac{1}{h\sqrt{2\pi}} \exp\left(\frac{-(x - x_i)^2}{2h^2}\right), \tag{2}$$

where $h$ is known as the bandwidth, controlling the kernel's width and thus the smoothness of the resulting density estimate. By summing the contributions of these kernels and normalizing by the total number of data points $n$, we obtain the estimated density at any point $x$:

$$\hat{f}(x) = \frac{1}{n}\sum_{i=1}^{n} K_h(x - x_i). \tag{3}$$

Each Gaussian kernel has a standard deviation equal to the bandwidth $h$. Consequently, the region within approximately $\pm h$ of each data point $x_i$ corresponds to the $1\sigma$ interval, containing roughly 68% of the kernel's mass. Similarly, the interval of approximately $\pm 2h$ for each point corresponds to about 95% of the kernel's mass, analogous to a $2\sigma$ confidence region in parametric Gaussian terms. Therefore, the KDE curves inherently represent the aggregation of these $1\sigma$ regions from each data point, visually illustrating where data points most densely cluster. Results of the burst detections for the complete time interval and their distribution are presented in Figure 4. In the top left panel, semitransparent histograms (normalized to density) and overlaid KDE curves compare the amplitude distributions for the Northern (blue) and Southern (orange) Hemispheres. Both hemispheres produce bursts spanning ≈10–180 SSN, but their central tendencies differ: the Northern Hemisphere amplitudes cluster more tightly ≈40 SSN, giving a single, well-defined mode, whereas the Southern Hemisphere amplitudes peak slightly higher between ≈60 and 90 SSN and exhibit a broader right-hand tail. This indicates that while typical burst strengths are similar in both hemispheres, very large bursts with amplitudes above ≈100 SSN occur more frequently in the Southern Hemisphere. These burst amplitude properties play a vital role in determining the burst amplitudes during the test and forecast periods. The top right panel of Figure 4 shows the corresponding burst-duration distributions in days. Here the Northern Hemisphere (blue) displays a wider, somewhat bimodal shape with a significant density ≈ 200–300 days and a secondary plateau near ≈350–400 days, extending out to ≈600 days. Southern Hemisphere durations (orange) are more narrowly concentrated at ≈200–250 days and drop off more steeply beyond 400 days. Thus, although both hemispheres share comparable median lifetimes, the Northern Hemisphere produces a larger fraction of very long-lived bursts, whereas the Southern Hemisphere's bursts tend to be more uniform in duration. Such a range of amplitudes and durations of the bursts is predicted by the theoretical models (M. Dikpati et al. 2018a), which discuss seasons during the SC.

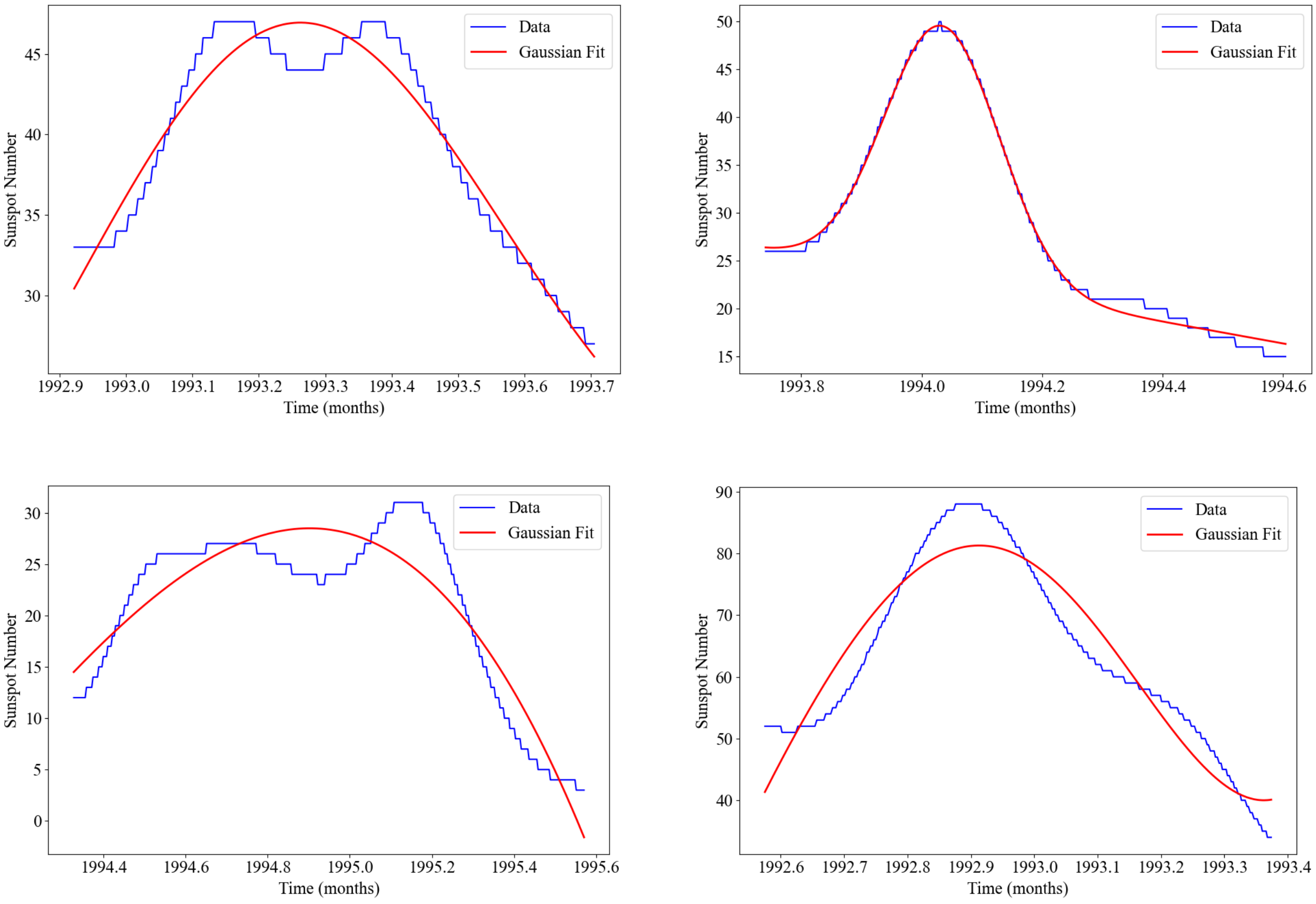

**Figure 3.** Example of Gaussian fitting to bursts observed in the time series. The blue line shows the 4 month smoothed SSN data, and the red line shows the fitted Gaussian function plus a linear background.

Burst properties during the SC's rising, maximum, and declining phases are visualized with violin plots derived from the KDEs. In each violin plot, the KDE curve is symmetrically mirrored about a central vertical axis. At each data point along the axis, the width of the violin is scaled proportionally to the estimated local density, visually emphasizing regions where observations are densely clustered (represented by wider sections) versus regions where data are sparse (represented by narrower sections). Thus, the characteristic "bulge" in the violin plot directly reflects the underlying empirical data distribution (J. L. Hintze & R. D. Nelson 1998). The violin plot widths involve the kernel's standard deviation $h$. Each Gaussian kernel encompasses ≈68% of its probability mass within the range $x_i \pm h$ (the so-called $\pm 1\sigma$ interval) and ≈95% within the range $x_i \pm 2h$ ($\pm 2\sigma$ interval). Wider regions of the violin plots reflect intervals where the majority of observations cluster tightly, whereas narrower regions denote less densely populated intervals. The bottom-left panel of Figure 4 presents the distribution of burst durations for the Northern (blue) and Southern (orange) Hemispheres. There were no bursts detected during the minimum phase of the SC. During the rising phase of the SC, the amplitude distributions for both hemispheres appear relatively symmetric, with densities peaking in the range of 50 and 100 SSN. The Northern Hemisphere shows two ranges related to peak density at ≈40 SSN and ≈120 SSN. The Southern Hemisphere shows that most bursts have an amplitude of ≈50 SSN, but shows more bursts of ≈150 SSN. In contrast, during the cycle maximum, the amplitude distributions reveal significant differences between hemispheres. The Northern Hemisphere distribution prominently peaks between 75 and 150 SSN, forming an asymmetrical shape with a noticeable extension toward higher amplitudes up to ≈200 SSN. This extended upper tail indicates that the Southern Hemisphere experiences bursts of notably higher amplitudes during this phase, albeit less frequently. The declining phase sees a convergence in distribution profiles between the two hemispheres. The Northern Hemisphere's distribution demonstrates a clear, symmetrical Gaussian-like shape centered ≈50–100 SSN, pointing to a stable pattern of moderate burst amplitudes during the solar decline. Similarly, the Southern Hemisphere displays a comparable Gaussian-like symmetry, with amplitude densities concentrated primarily within 50–90 SSN, albeit with a slightly narrower spread than observed in the rising phase.

The bottom right panel of Figure 4 compares the burst-duration distribution across three distinct SC phases between the Northern and Southern Hemispheres. During the rising phase, bursts in the Northern Hemisphere exhibit a bimodal distribution characterized by two prominent peaks, roughly centered at durations of ≈150 and 200 days. This bimodality implies two distinct populations or physical regimes of bursts: short-duration events of half a year and a a longer-duration group that extends toward ≈250–400 days. In contrast, the Southern Hemisphere during this rising phase displays a smoother, approximately Gaussian-like distribution, peaking primarily at about 200 days and tapering symmetrically on either side. This smoother shape indicates a more uniform burst-duration pattern in the Southern Hemisphere without strong subgroup distinctions. At solar maximum, both hemispheres tend to exhibit distributions with considerable width, indicating a wide spread of burst durations. The Northern Hemisphere distribution at solar maximum has a dominant peak between ≈200 and 350 days, accompanied by an extended tail reaching durations near 700 days. This tail highlights the presence of rare but exceptionally long bursts that occur even at the cycle's peak activity period. Similarly, the Southern Hemisphere shows a smooth distribution centered broadly at ≈250–350 days, suggesting a relatively uniform set of durations with occasional prolonged events visible in the distribution's moderate upper tail. In the declining phase, distributions in both hemispheres again widen, highlighting increased variability in burst durations. The Northern Hemisphere distribution is again prominently bimodal, with one cluster of durations concentrated near 100 days and another at ≈300 days. Meanwhile, the Southern Hemisphere maintains its approximately Gaussian-shaped profile, centered at ≈200 days; a small fraction of bursts reaching durations around 700 days.

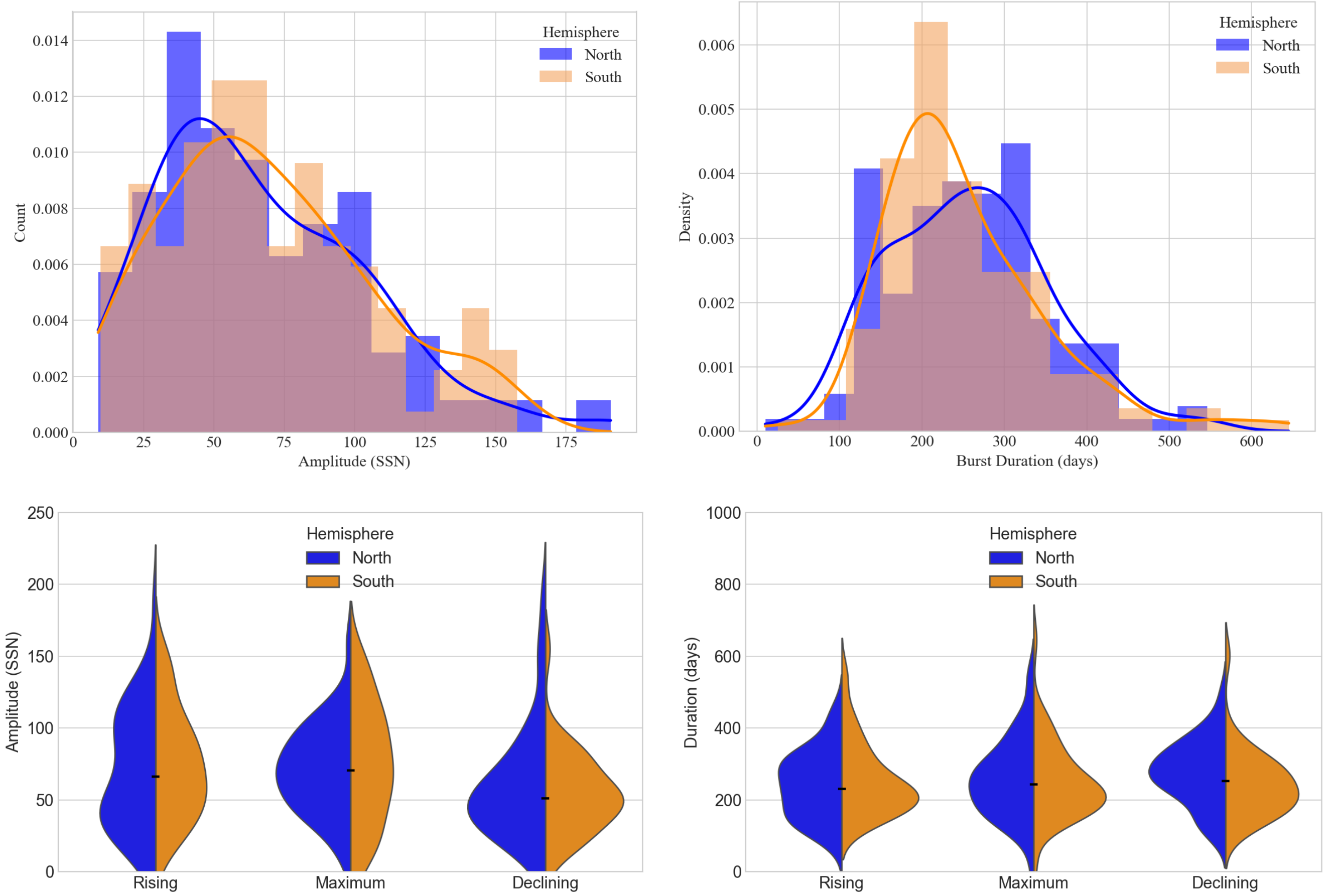

**Figure 4.** Comparison of solar burst properties between the Northern and Southern Hemispheres. (a) Distribution of amplitudes for Northern and Southern Hemispheres bursts. (b) Distribution of durations (in months) for Southern Hemisphere bursts. (c) Violin plot comparison of amplitudes relative to SC phase and hemisphere. (d) Violin plot comparison of duration to SC phase and hemisphere. All panels aggregate bursts over SC12–SC25 (1878–2025).

## 3. Using SARIMA to Statistically Study Smoothed Sunspot Number Trends

To forecast the occurrence of upcoming burst events in each hemisphere, we adopt two complementary modeling strategies tailored specifically to the Northern and Southern Hemisphere sunspot time series. An SARIMA (G. Box & G. Jenkins 1976; R. J. Hyndman & G. Athanasopoulos 2018) model is employed to accurately capture the underlying regular cyclic and seasonal patterns observed in the monthly SSN and their burst properties. SARIMA can distinguish between the cyclic behavior of the SC, which is a long-term predictable model, and dynamic bursts, which are a part of its differencing strategy (R. J. Hyndman & G. Athanasopoulos 2018).

The SARIMA model is formally expressed as

$$\phi_p(B)\Phi_P(B^m)(1 - B)^d(1 - B^m)^D y_t = \theta_q(B)\Theta_Q(B^m)\epsilon_t. \quad (4)$$

Here, $y_t$ represents the observed SSN at month $t$, $B$ is the backshift operator ($By_t = y_{t-1}$), $m$ denotes the seasonal periodicity, $d$ and $D$ are nonseasonal and seasonal differencing orders, respectively, and $\epsilon_t$ is Gaussian white noise with a mean of zero and variance $\sigma^2$. Polynomials $\phi_p(B)$, $\Phi_P(B^m)$, $\theta_q(B)$, and $\Theta_Q(B^m)$ denote autoregressive and moving-average components at both seasonal and nonseasonal levels (G. Box & G. Jenkins 1976; R. J. Hyndman & G. Athanasopoulos 2018). These components collectively define the SARIMA model, which has been utilized for sunspot forecasting in recent work (Q. Xu et al. 2024).

The SARIMA modeling process begins with the monthly averaged SSNs spanning from 1878 to 2024. To ensure accuracy and reliability, each hemisphere is analyzed individually, acknowledging their distinct solar activity patterns. The fixed SARIMA model parameters were chosen separately for each hemisphere based on historical performance and statistical validation criteria. For each hemisphere, we divide the data into training, test, and forecast. The top panel of Figure 5 presents the Northern Hemisphere SSNs from 1878 to 2025, with the dotted line denoting the historical segment used for model fitting, the solid line from 2024 July to 2025 January showing the test interval, and the continuation from 2025 January to 2026 April giving the hindcast and the forecast. The lower panel shows the analogous Southern Hemisphere series: the dotted curve is the training data up to 2024 October, while the solid curve covers the test window (2024 November–2025 January) and the subsequent hindcast with forecast. The bottom two panels of Figure 5 show the forecasted monthly SSNs for SC25 in the Northern (top) and Southern (bottom) Hemispheres, zoomed into 2026 March. The solid blue and orange lines represent the forecasts for the Northern and Southern Hemispheres, respectively. The vertical dashed lines indicate key epochs: the official start of SC25 (2019 December, gray), the beginning of the test interval (2024 July for the Northern Hemisphere and 2024 November for the Southern Hemisphere, shown as blue and orange dashed lines, respectively), and the onset of the forecast period (2025 February, green). The Northern Hemisphere forecast shows moderate burst-like activity with peaks around 60–90 SSN, while the Southern Hemisphere is predicted to experience higher-amplitude bursts, peaking above 140 SSN before beginning a gradual decline. These hemispheric asymmetries highlight the dynamic and distinct evolution of solar activity in each hemisphere during the ascending phase of SC25.

Model parameters were chosen with `pmdarima`'s `auto_arima`.[5] The SARIMA model orders for each hemisphere were selected to minimize the Akaike information criterion (AIC), resulting in values of 15,405.70 for the Northern Hemisphere and 15,476.56 for the Southern Hemisphere. This criterion ensures a good model fit while penalizing unnecessary complexity, thereby reducing the risk of overfitting. Following the model fit, the predicted values were smoothed using a centered rolling average.

To select an appropriate window size, a sensitivity analysis is performed for windows $m \in \{2, 3, 4, 5, 6\}$ months. The impact of each window is evaluated based on its out-of-sample forecast accuracy (mean absolute error (MAE) and root mean squared error (RMSE)) and its effect on the stability of the forecast, measured by the number of detected SC peaks (Table 1). MAE is defined as

$$\mathrm{MAE} = \frac{1}{n}\sum_{i=1}^{n} | y_i - \hat{y}_i |, \tag{5}$$

where $y_i$ and $\hat{y}_i$ are the observed and predicted values, respectively, and $n$ is the number of time steps in the test set. MAE reports the average deviation of the forecast from the truth, expressed in the same physical units as the data, so smaller values indicate more accurate predictions.

A complementary metric is the RMSE, which is particularly sensitive to large forecast errors. RMSE is defined as

$$\mathrm{RMSE} = \sqrt{\frac{1}{n}\sum_{i=1}^{n} ( y_i - \hat{y}_i )^2}, \tag{6}$$

where $y_i$ and $\hat{y}_i$ are the observed and predicted values, respectively, and $n$ is the number of time steps in the test set. By squaring the difference between the observed and predicted values, RMSE places a much higher penalty on large errors than on small ones. A model with a low RMSE is therefore not only accurate on average but is also more stable and reliable, as it demonstrates an ability to avoid significant, outlier forecast failures.

The impact of each window is evaluated based on its out-of-sample forecast accuracy (AIC, MAE, and RMSE) and its effect on the stability of the forecast, measured by the number of detected SC peaks (Table 1). The analysis reveals a clear trade-off between forecast accuracy and stability. A 4 month window is selected as it represents the optimal compromise. For the Northern Hemisphere, this window minimizes the MAE and RMSE at 18.04 and 18.47, respectively, while producing a stable forecast with only two peaks. For the Southern Hemisphere, a 4 month window provides a robust forecast with a low RMSE of 8.83, avoiding the higher RMSE and greater number of spurious peaks associated with larger windows. This choice ensures that the final forecast is robust against short-term noise without distorting the underlying structure of the SC. To account for the anticipated decline in solar activity following the projected solar maximum approximately 2024 December (L. A. Upton & D. H. Hathaway 2023), a multiplicative decay factor was applied, reducing the amplitude by 20% every 12 months.

### 3.1. Secondary Bursts

Secondary bursts in solar activity can be effectively represented by superimposing additional Gaussian profiles onto the primary burst Gaussian curve. Numerically, the complete burst profile with a primary and a secondary burst component can be expressed explicitly as the sum of two second-order Gaussian functions:

$$y(t) = A_1 \exp\left(-\frac{(t - t_1)^2}{2\sigma_1^2}\right) + A_2 \exp\left(-\frac{(t - t_2)^2}{2\sigma_2^2}\right), \tag{7}$$

where $y(t)$ is the total SSN at time $t$. $A_1$ and $A_2$ represent the peak amplitudes of the primary and secondary bursts, respectively (maximum SSN). $t_1$ and $t_2$ denote the peak timings (central points) of the primary and secondary Gaussian bursts, respectively. $\sigma_1$ and $\sigma_2$ characterize the duration or width of each burst. Smaller values correspond to sharper bursts, and larger values indicate broader bursts.

This numerical expression represents the sum of two Gaussian terms, each a second-order function due to the squared exponent that captures independent solar burst events that may overlap or occur in close temporal proximity. This additional criterion is incorporated into the SARIMA model, thereby enhancing its capability to replicate the complex burst shapes observed in actual solar activity data. Historical data inform the timing and amplitude of the forthcoming burst, and this method parallels successful parametric forecasting approaches in solar physics (H. Zhu et al. 2022; Q. Xu et al. 2024).

### 3.2. SARIMA Parameter Testing for the Northern Hemisphere

To determine the optimal SARIMA configuration for forecasting sunspot activity in the Northern Hemisphere, a comprehensive parameter-testing pipeline is developed and executed. This involved 288 tests with different combinations of SARIMA parameters. This routine systematically explores a range of candidate seasonal periods $m$ and differencing parameters ($d$ and $D$), while holding the maximum model orders fixed at $(p, q, P, Q) \leqslant (5, 5, 2, 2)$. The values for $(p, q, P, Q)$ are restricted to be $\leqslant(5, 5, 2, 2)$, due to avoiding overfitting the SC trends, as per arguments in R. J. Hyndman & G. Athanasopoulos (2018). For each configuration, an SARIMA model is trained using monthly averaged sunspot data from 1878 to 2025, and its forecasting accuracy is evaluated over a fixed test window from 2024 June to 2025 January.

[5] https://alkaline-ml.com/pmdarima/modules/generated/pmdarima.arima.auto_arima.html

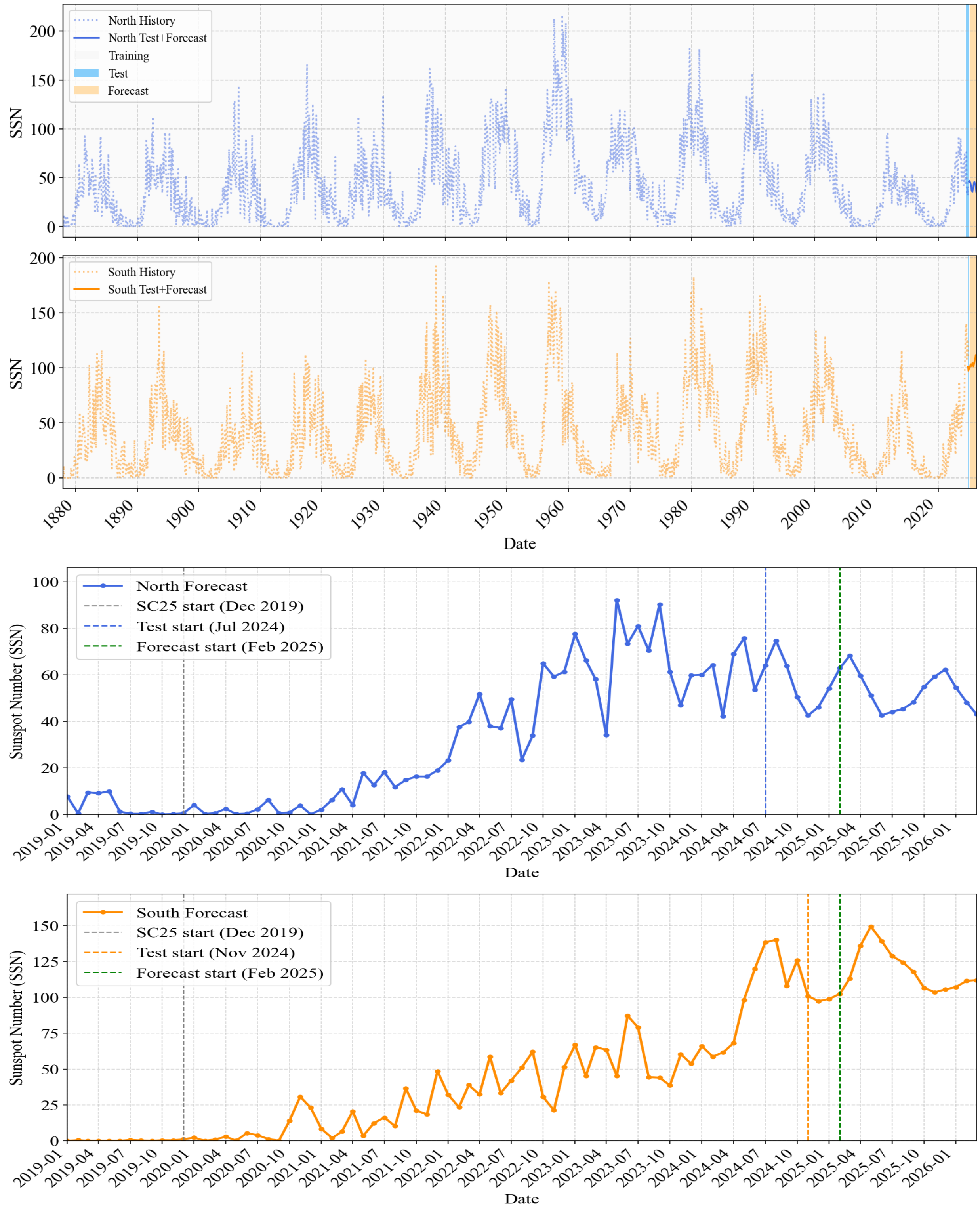

**Figure 5.** Top panel: Northern Hemisphere monthly SSN, from 1878–2026. Second panel from the top: Southern Hemisphere monthly SSN, from 1878–2026. Third panel from the top: forecasted monthly SSN for SC25, Northern Hemisphere, zoomed to 2026 March. Bottom panel: forecasted monthly SSN for SC25, Southern Hemisphere, zoomed to 2026 March. In the top two panels, dotted lines indicate training data, solid lines denote the hindcast test intervals (Northern Hemisphere: 2024 July–2025 January; Southern Hemisphere: 2024 November–2025 January); subsequent segments show the forecast. Background shading marks the training (gray), test (blue), and forecast (tan) periods. Axes are SSN ($y$) vs. calendar year ($x$). In the third and bottom panels, the solid blue and orange curves are the Northern and Southern Hemisphere forecasts, respectively. Vertical dashed lines mark key epochs: SC25 start (2019 December, gray), test-interval start (Northern Hemisphere: 2024 July; Southern Hemisphere: 2024 November, blue and orange, respectively), and forecast start (2025 February, green).

**Table 1**
Sensitivity of the SARIMA Baseline Forecast to the Monthly Smoothing Window Size (m)

| Months | Hemisphere | Forecast Peak Count | Forecast Peak Months (YYYY-MM) | AIC | MAE | RMSE |
|---|---|---|---|---|---|---|
| 2 | North | 3 | 2024-09, 2025-02, 2025-11 | 15,405.70 | 18.78 | 19.54 |
| 2 | South | 4 | 2025-03, 2025-07, 2025-10, 2026-01 | 15,476.56 | 15.23 | 15.24 |
| 3 | North | 4 | 2024-10, 2025-02, 2025-07, 2025-11 | 15,405.70 | 18.91 | 19.12 |
| 3 | South | 3 | 2025-03, 2025-08, 2026-02 | 15,476.56 | 8.43 | 8.83 |
| 4 | North | 2 | 2025-03, 2025-12 | 15,405.70 | 18.04 | 18.47 |
| 4 | South | 3 | 2025-05, 2025-08, 2026-03 | 15,476.56 | 8.43 | 8.83 |
| 5 | North | 2 | 2025-02, 2025-12 | 15,405.70 | 20.07 | 20.66 |
| 5 | South | 5 | 2025-01, 2025-05, 2025-08, 2025-11 | 15,476.56 | 7.11 | 9.18 |
| … | … | … | 2026-03 | … | … | … |
| 6 | North | 1 | 2025-02 | 15,405.70 | 19.73 | 20.27 |
| 6 | South | 3 | 2025-01, 2025-05, 2025-08 | 15,476.56 | 7.11 | 9.18 |

For the Northern Hemisphere, the optimal configuration is found to be an SARIMA (3, 0, 5) × $(2, 1, 0)_9$ model. This setting is represented in the top row in Table 2. This setup, with nonseasonal autoregressive order $p = 3$, moving-average order $q = 5$, seasonal autoregressive order $P = 2$, seasonal moving-average order $Q = 0$, seasonal period $m = 9$ months, and seasonal differencing $D = 1$, achieves the lowest AIC ≈ 15,397 among the tested configurations. It also produces the best forecast accuracy, with an MAE of ≈11.98 and an RMSE of ≈16.72. These metrics indicate that the model consistently tracked the observed SSNs with the smallest average and largest-error deviations compared to alternative configurations, many of which showed MAE $\gtrsim$ 19 and RMSE $\gtrsim$ 21. Physically, the 9 month seasonal period reflects a quasi-annual modulation in northern sunspot activity. The relatively high moving-average order ($q = 5$) in the nonseasonal component suggests that short-term irregular fluctuations with month-to-month disturbances in sunspot counts are more prominent in the north and require a longer memory of past forecast errors to be modeled effectively. The inclusion of $P = 2$ and $D = 1$ indicates that variations over the last ∼1–2 seasonal cycles influence current activity, and that a single step of seasonal differencing is necessary to stabilize long-term seasonal trends. We then visually inspect zoomed-in plots to identify model behavior that best captures the observed burst timing and amplitude characteristics. Some of the settings for the Northern Hemisphere are mentioned in Table 2. Through this process, the optimal model for the Northern Hemisphere was selected as SARIMA (3, 0, 5) × $(2, 1, 0)_9$.

*3.3. SARIMA Parameter Testing for the Southern Hemisphere*

To identify the optimal SARIMA model configuration for forecasting Southern Hemisphere sunspot activity, a parallel pipeline has been executed using the same methodology as for the Northern Hemisphere. Monthly averaged sunspot data from 1878 through 2024 have been used to train the models, with forecasts evaluated over a common test interval from 2024 January to 2025 January. In the final configuration, we use the test period from 2024 October to 2025 January. The SARIMA test suite systematically scans combinations of seasonal periods $m$ and differencing orders ($d$ and $D$), while fixing the maximum model orders at $(p, q, P, Q) \leqslant (5, 5, 2, 2)$. Evaluation is based on the MAE between the model forecast and the observed SSNs over the test interval. In addition, each configuration's performance is visually assessed to identify model behavior that most accurately reproduces the timing and amplitude of known burst events.

**Table 2**
Top 10 SARIMA Parameter Test Results for the Northern Hemisphere

| $m$ | $d$ | $D$ | $p$ | $q$ | $P$ | $Q$ | $s$ | AIC | MAE | RMSE |
|---|---|---|---|---|---|---|---|---|---|---|
| 9 | 0 | 1 | 3 | 5 | 2 | 0 | 9 | 15,397.07 | 11.98 | 16.72 |
| 9 | 0 | 1 | 3 | 4 | 2 | 0 | 9 | 15,780.55 | 19.41 | 21.99 |
| 9 | 0 | 0 | 5 | 4 | 2 | 0 | 9 | 15,403.77 | 19.36 | 21.42 |
| 10 | 0 | 1 | 3 | 1 | 2 | 0 | 10 | 15,840.76 | 19.36 | 21.41 |
| 10 | 1 | 1 | 5 | 1 | 2 | 0 | 10 | 15,860.75 | 19.44 | 22.50 |
| 8 | 0 | 1 | 3 | 1 | 2 | 0 | 8 | 15,858.58 | 19.37 | 21.56 |
| 8 | 1 | 1 | 5 | 1 | 2 | 0 | 8 | 15,884.85 | 19.44 | 22.60 |
| 11 | 0 | 1 | 4 | 4 | 2 | 0 | 11 | 15,822.56 | 19.61 | 22.86 |
| 11 | 1 | 0 | 2 | 1 | 0 | 0 | 11 | 15,441.88 | 19.35 | 21.36 |
| 12 | 1 | 0 | 2 | 2 | 1 | 1 | 12 | 15,421.73 | 19.34 | 21.34 |

**Note.** With the top setup used for the final model.

In the Southern Hemisphere, the best-performing configuration is an SARIMA (3, 0, 1) × $(2, 1, 0)_{10}$ model (Table 3). This model features nonseasonal autoregressive order $p = 3$, moving-average order $q = 1$, seasonal autoregressive order $P = 2$, seasonal moving-average order $Q = 0$, seasonal period $m = 10$ months, and seasonal differencing $D = 1$. It achieves an AIC of ≈15,476, with an MAE of ≈16.36 and an exceptionally low RMSE of ≈16.64, making it highly resistant to large, outlier forecast errors despite a slightly higher average error than the Northern Hemisphere model. The 10 month seasonal period points to a subtly longer recurrence time for intermediate-scale activity bursts in the south compared to the north. The lower moving-average order ($q = 1$) indicates that short-term variability is less complex, and the southern sunspot series can be well described using a shorter memory of past forecast errors. Interestingly, several alternative southern configurations with much longer seasonal periods (18–60 months) produce similar AIC and MAE values, suggesting that southern activity may be influenced by a broader mix of periodicities and lacks a sharply defined single seasonal timescale. As in the north, the values $P = 2$ and $D = 1$

**Table 3**
Top 10 SARIMA Parameter Test Results for the Southern Hemisphere

| $m$ | $d$ | $D$ | $p$ | $q$ | $P$ | $Q$ | $s$ | AIC | MAE | RMSE |
|---|---|---|---|---|---|---|---|---|---|---|
| 10 | 0 | 1 | 3 | 1 | 2 | 0 | 10 | 15,476.56 | 16.36 | 16.64 |
| 9 | 0 | 0 | 5 | 1 | 2 | 0 | 9 | 15,315.48 | 16.90 | 23.60 |
| 10 | 0 | 1 | 5 | 1 | 2 | 0 | 10 | 15,747.23 | 19.75 | 27.17 |
| 11 | 0 | 0 | 5 | 1 | 2 | 0 | 11 | 15,330.76 | 16.70 | 23.52 |
| 8 | 1 | 0 | 2 | 2 | 0 | 0 | 8 | 15,322.36 | 18.26 | 25.79 |
| 12 | 1 | 0 | 2 | 2 | 0 | 0 | 12 | 15,322.36 | 18.26 | 25.79 |
| 18 | 1 | 0 | 2 | 1 | 0 | 1 | 18 | 15,326.29 | 18.24 | 26.05 |
| 24 | 1 | 0 | 2 | 2 | 0 | 0 | 24 | 15,322.36 | 18.26 | 25.79 |
| 48 | 1 | 0 | 2 | 2 | 0 | 0 | 48 | 15,322.36 | 18.26 | 25.79 |
| 60 | 1 | 0 | 2 | 2 | 1 | 0 | 60 | 15,320.10 | 18.05 | 25.38 |
| 10 | 0 | 2 | 3 | 1 | 2 | 2 | 10 | 16,695.43 | 14.75 | 22.49 |

**Note.** With the top setup used for final model.

highlight the role of the preceding 1–2 seasonal cycles and the need to remove residual seasonal nonstationarity. Overall, the Southern Hemisphere model reflects a smoother error structure and a slightly longer dominant cycle, consistent with the historical tendency for the Southern Hemisphere to produce fewer but often more intense bursts. Table 3 summarizes a selection of top-performing SARIMA configurations.

The optimal model for the Southern Hemisphere is determined to be SARIMA $(3, 0, 1) \times (2, 1, 0)_{10}$. This configuration includes three autoregressive (AR) terms and one moving average (MA) term in the nonseasonal component, with no nonseasonal differencing, and a seasonal cycle length of 10 months. The selected model shown in the top row of Table 3.

## 4. Machine Learning Random Forest Regression for Burst Amplitude and Duration from Gaussian Fits

Forecasting the amplitude and duration of solar activity is the final step in the forecasting model. Properties of bursts are treated as a supervised regression problem. This is addressed using RF regression, selected for its robustness, ability to handle nonlinear relationships, and resistance to overfitting (L. Breiman 2001).

RFs are ensemble learning methods that average predictions from multiple decision trees to improve generalization and reduce variance. Given training data $\{(x_i, y_i)\}_{i=1}^{n}$, the predicted value for a new input $x$ is obtained by

$$\hat{f}(x) = \frac{1}{M}\sum_{m=1}^{M} T_m(x), \tag{8}$$

where $T_m(x)$ is the prediction made by the $m$th tree in a forest of $M$ total trees.

We implement the RF with `scikit-learn`'s `RandomForestRegressor` using 200 trees grown on bootstrap samples. The split criterion is the default squared error. To limit overfitting on the small burst catalog we require at least two samples per leaf and all other depth controls are left at the defaults (effectively unlimited depth, and no limit on leaf nodes). For regression, we enable out-of-bag (OOB) estimation and report the OOB $R^2$ as an internal generalization check. The input vector for each burst consists of the hemisphere indicator and lagged monthly hemispheric SSNs, and the target is the burst amplitude taken from the Gaussian fit; the burst duration is modeled and evaluated separately. Model selection

**Table 4**
Random Forest Model Performance via Five-fold Cross Validation

| Hemisphere | MAE (SSN) | RMSE (SSN) | $R^2$ Score |
|---|---|---|---|
| North | 13.26 | 20.15 | 0.706 |
| South | 11.91 | 16.50 | 0.794 |
| Overall | 12.60 | 18.46 | 0.748 |

**Note.** Metrics shown are the MAE, RMSE, and the coefficient of determination ($R^2$).

is kept fixed and performance is assessed using MAE, RMSE, and $R^2$.

For each burst, a set of predictor features is extracted from the time series preceding the burst onset. These features include lagged monthly SSNs at 1–5 months before the burst, as well as the hemisphere indicator. The feature vector is thus

$$X = [\mathrm{SSN}_{t-1}, \mathrm{SSN}_{t-2}, \ldots \mathrm{SSN}_{t-5}, \text{Hemisphere}], \tag{9}$$

where $\mathrm{SSN}_{t-k}$ denotes the sunspot number $k$ months prior to the burst peak. This lagged structure allows the model to capture precursor patterns in solar activity that typically precede a burst.

An RF model is trained on the combined historical burst catalog from both hemispheres to predict burst amplitude based on precursor sunspot activity. The hemisphere itself is included as a feature to allow the model to learn distinct behaviors (E. Gurgenashvili et al. 2017). The model's performance is evaluated using MAE, RMSE, and the $R^2$ score.

The model demonstrates strong predictive performance, as detailed in Table 4. Overall, the model achieves an $R^2$ score of 0.748, indicating that it explains approximately 75% of the variance in burst amplitudes. The cross-validated MAE for amplitude prediction is 12.60 SSN, a significant improvement over previous model iterations. The duration model, evaluated separately, yields an MAE of about 1.8 months.

## 5. Final Forecasts: Combined SARIMA and Machine Learning Burst Injection

The final forecast is a hybrid model: SARIMA provides the baseline and an RF supplies burst amplitudes. For the Northern Hemisphere, the specification is SARIMA $(3, 0, 5) \times (2, 1, 0)_9$ (see Section 3.2); for the Southern Hemisphere, it is SARIMA $(3, 0, 1) \times (2, 1, 0)_{10}$ (see Section 3.3). Burst amplitudes are learned from the hemisphere indicator and lagged monthly hemispheric SSNs and injected as Gaussian components onto the SARIMA baseline (see Section 4). Figure 6 illustrates the final hemispheric forecasts, which integrate the SARIMA baseline with our machine learning–based burst prediction model. To provide a comprehensive assessment of the forecast's reliability and to address the distinct sources of uncertainty, the figure has been enhanced to include a full characterization of the model's predictive confidence. The plot distinguishes the historical training data (dotted line), the out-of-sample test data (dashed line), and the forecast period (solid line). A principal enhancement is the inclusion of a light shaded region representing the 95% confidence interval (CI) for the entire SARIMA baseline forecast, which directly quantifies the statistical uncertainty of the underlying trend. For individual forecasted bursts, the uncertainty is characterized in two complementary ways: vertical error bars (burst amplitude ± error) represent the historical standard deviation of burst amplitudes, while the horizontal shaded regions serve as a proxy for the burst's timing and duration, spanning 50% of the rise-to-decay period. The forecasts are presented across three panels: the Northern Hemisphere (top), the Southern Hemisphere (middle), and their combined total (bottom), showing monthly SSN predictions from 2024 through mid-2026. The vertical error bars and horizontal shaded regions shown for each forecast burst in Figure 6 constitute a pragmatic, data-driven proxy for the uncertainty (see Appendix B).

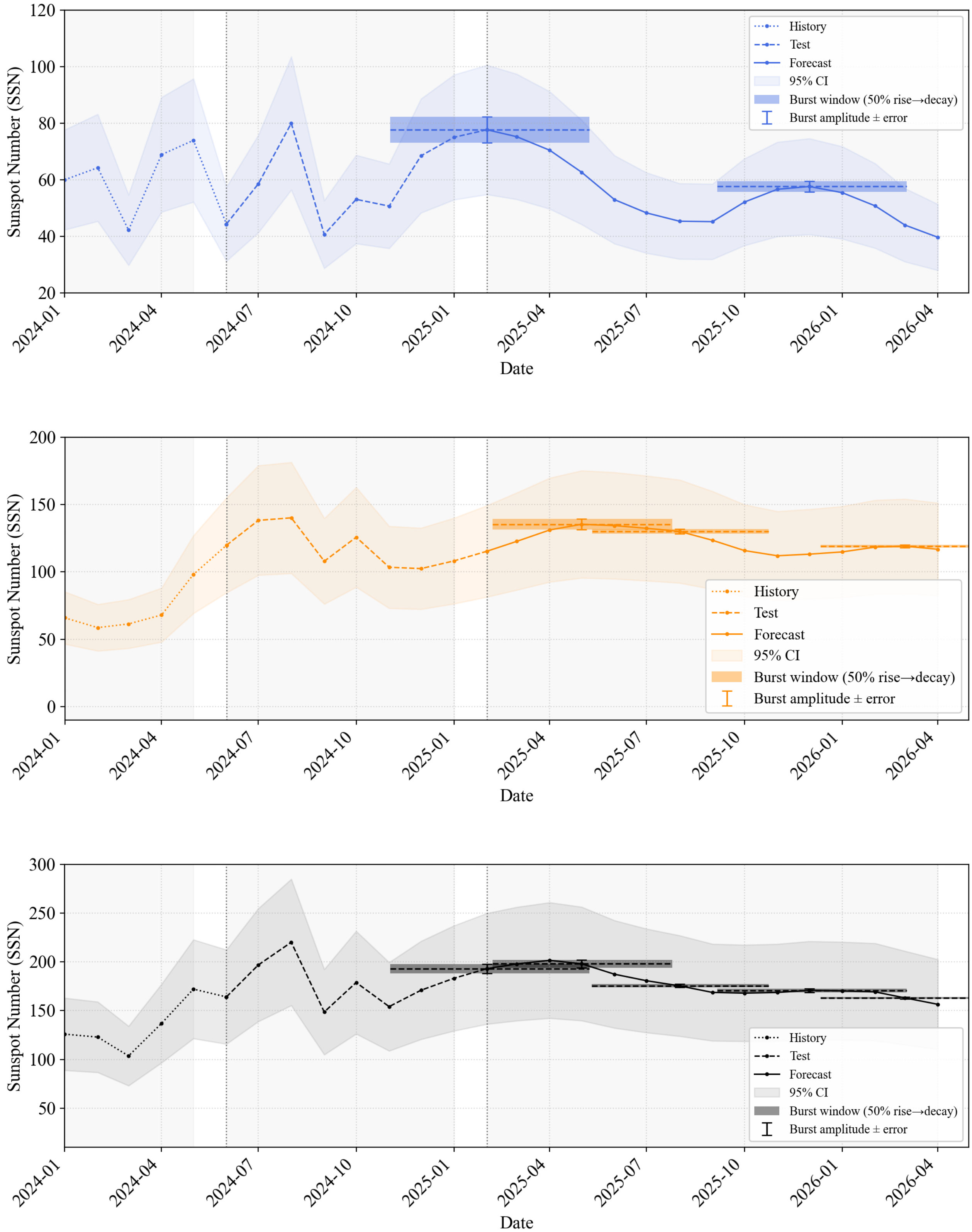

**Figure 6.** Zoomed-in burst-envelope forecasts from 2024 January to 2026 April. Top: the Northern Hemisphere SSN with inferred RF burst properties. Middle: the Southern Hemisphere counterpart. Bottom: total (Northern Hemisphere + Southern Hemisphere) SSN, highlighting the same burst timings and amplitudes.

These baselines are initially scaled to 80% of their original values (a 20% reduction) to account for a declining SC phase using an exponential decay with a 12 month half-life. This tapering approach is designed to mimic the typical postmaximum behavior of SCs and is supported by prior studies (e.g., L. A. Upton & D. H. Hathaway 2023). Burst amplitudes are predicted using RF models trained on catalogs of historical bursts that are characterized by Gaussian fits. These models incorporate lagged SSNs and hemispheric indicators as predictive features. Finally, forecast bursts are injected onto the processed baselines using asymmetric Gaussian envelopes. The parameters of these envelopes, specifically their amplitude, rise duration, and decay duration, are dynamically scaled based on the RF machine learning models.

In the Northern Hemisphere (top panel of Figure 6), the model identifies an initial burst peaking around 2025 March. This event, which functions as the hindcast (test) result for this study, exhibits a projected amplitude between 70 and 80 SSN and an approximate 9 month duration commencing in 2024 October. Furthermore, the model indicates no significant secondary burst activity during this specific hindcast period. Subsequently, a distinct forecasted burst, derived from the hybrid methodology, is projected to peak around 2025 December. This latter burst is anticipated to commence around 2025 October, conclude by 2026 March, and exhibit an amplitude between 55 and 65 SSN. Its lower projected amplitude, when compared to the hindcast burst, is attributed to model indications that the SC will be in its declining phase, having passed its peak.

In contrast, for the Southern Hemisphere (middle panel, Figure 6), the model delineates a primary forecast burst event. The activity associated with this burst commences in late 2024, influencing the hindcast result, particularly within the validation window of 2024 November to 2025 January. This primary event is projected to reach its main peak in early 2025 (around March–April, as indicated in the figure). This primary event initially presents as a structure with multiple peaks. To characterize its dominant component, the RF element of the hybrid model fits a parameterized Gaussian profile centered around the 2025 March peak. This representative Gaussian fit indicates a peak amplitude for the primary burst of between 130 and 140 SSN, with the overall activity from this event projected to extend until mid-2025. During the declining phase of this primary burst, the model also identifies a subsequent, smaller burst. The emergence of this secondary feature is primarily attributed to the SARIMA component's sensitivity to underlying cyclical patterns from adjacent SCs and expects the secondary burst. Finally, extending further into the forecast horizon, the model predicts the occurrence of a third burst around 2026 January, with an amplitude of around 120 SSN and lasting from 2025 December to 2026 April.

Collectively, these forecasts showcase a robust hybrid framework adept at capturing seasonal and burst-driven modulations in sunspot activity, while simultaneously offering valuable insight into near-term space-weather conditions and hemispheric asymmetries in the SC. The bottom panel of Figure 6 displays the combined total SSN forecast, which represents the sum of the previously detailed Northern and Southern Hemisphere predictions. Consistent with the individual hemisphere plots, distinct line styles are used: a dotted line for the historical training data, a dashed line for the test/validation period, and a solid line with markers for the forecast period. The historical data leading into the forecast show significant variability. The test phase, for instance, is characterized by a sharp, prominent peak where the total SSN reached ≈220 around 2024 August–September. Following this, the forecast period commences in early 2025. The model projects a primary peak in the total SSN of around 205–210 occurring in 2025 March–April. For key forecasted burst events such as this one, vertical error bars are included to represent the standard deviation of historical burst amplitudes, while the accompanying horizontal shaded regions denote the typical range or standard deviation of historical burst durations. After this primary forecasted peak, the total SSN is predicted to decline to ≈170 by mid-2025. Subsequently, the model indicates a secondary, broader period of heightened activity where the total SSN is expected to plateau around 170–175 from late 2025 into early 2026. This phase is also marked with uncertainty indicators for burst amplitude and duration. Thereafter, the forecast projects a general decline in total sunspot activity through mid-2026.

## 6. Concluding Remarks

Solar magnetic activity, which causes hazardous space weather including a profound influence on our technological society, occurs on various timescales. Solar activity occurring on the well-known decadal timescale shows systematic variability with sinusoidal patterns, the intensity of which is measured by the amplitude of that sinusoid. A stronger cycle is considered to create more energetic events compared to weaker ones. On the other hand, daily variability of solar activity appears random, making it nearly impossible to predict what will happen the next day. Forecasting how many sunspots will emerge or how many will erupt as CMEs and major flares only adds to this challenge. This is very much like the earthquake predictions, where the timing, location and magnitude of the earthquakes are considered nearly impossible to predict. However, solar activity variability occurring on the intermediate timescales of weeks to months is not fully random; instead, these quasiperiodic bursts of enhanced activity, during which major space-weather "seasons" occur, have ample systematic variability to ensure predictions. Such predictions can be of high value to society because an adequate lead time of several weeks can be provided to help prepare for the impacts of the upcoming "season" of enhanced bursts of solar activity. This is the aim of this work.

There is growing motivation, both observational and theoretical, to explore the features of solar activity bursts and to understand why the decadal SC progresses through alternating bursty and quiet phases. The underlying mechanism driving short-term, quasiperiodic bursts in solar activity that is often termed the "seasons" of space weather, is attributed to global MHD instabilities operating within the Sun's tachocline. In particular, MHD shallow-water tachocline models (e.g., T. V. Zaqarashvili et al. 2007; M. Dikpati et al. 2017) suggest that these instabilities generate Rossby waves. The nonlinear interactions of these waves with magnetic fields and differential rotation are thought to produce TNOs with periods ranging from 2 to 20 months. For instance, T. V. Zaqarashvili et al. (2010) demonstrate that for a 10 kG magnetic field, unstable fast Rossby wave harmonics can exhibit periods of ≈5 months, with these periods varying according to the mode's symmetry. Similar processes have also been explored in the Sun's supergranulation layer, while dynamo models (e.g., F. Inceoglu et al. 2019) further indicate that the interplay between plasma flows and magnetic fields can generate such short-term variability.

Hence, we divide solar activity on monthly timescales, which is shaped by two intertwined processes: the first one is the smoothly varying magnetic cycle that gives rise to the familiar 11 yr modulation of the SSN, and the second a burst-like enhancement that emerges on quasi-annual and quasi-biennial timescales (D. H. Hathaway 2015). Although theoretical models offer a robust basis for the quasi-annual and quasi-biennial periodicities observed in solar activity, their direct predictive capability for burst events remains limited, primarily due to uncertainties in parameter estimation and the complexity of the underlying processes. Consequently, they often struggle to accurately forecast the precise timing, amplitude, and hemispheric sequence of individual burst events. Recent advancements in data assimilation and machine learning, however, present promising avenues to address these limitations. The current lack of reliable forecasting methods at intermediate lead times ranging from weeks to several months poses a critical gap for operational space-weather forecasting.

In this work, we present a hybrid forecasting framework that unites several key contributions. We couple seasonal autoregressive modeling with burst-specific Gaussian envelopes, whose parameters are supplied by an RF regressor, and integrate these into a statistical SARIMA baseline. Using this hybrid model, we predict both the amplitude and duration of upcoming bursts, thereby addressing a long-standing gap in intermediate lead time solar prediction. While traditional SC forecasts effectively track the decadal rise and fall of SSN, they rarely resolve the quasi-seasonal bursts with periods between 2 and 20 months that dominate space-weather risk on operational timescales. By explicitly modeling these bursts, we translate the physically motivated concept of TNOs into a tractable and data-driven forecasting workflow.

To accurately forecast hemispheric sunspot activity, we employ two separate SARIMA configurations. For the Northern Hemisphere we apply SARIMA $(3, 0, 5) \times (2, 1, 0)_9$, and SARIMA $(3, 0, 1) \times (2, 1, 0)_{10}$ in the Southern Hemisphere. The necessity of different SARIMA configurations in the Northern Hemisphere and the Southern Hemisphere highlights genuine dynamical contrasts that arise because the northern activity is more regular and therefore benefits from a higher nonseasonal MA order that captures finer short-term deviations. The Southern Hemisphere activity, by contrast, is punctuated by frequent, high-amplitude spikes that force the model toward a lower MA order and a slightly longer seasonal lag. These empirical choices align with physical expectations wherein the Southern Hemisphere has historically shown larger intermittency in magnetic flux emergence, which our burst catalog confirms (e.g., M. Temmer et al. 2006; A. A. Norton & J. C. Gallagher 2010; S. W. McIntosh & R. J. Leamon 2014).

We use an RF regression machine learning model to supply burst amplitudes and durations, allowing them to vary smoothly with precursor conditions, which include both lagged SSNs and hemispheric context. We use ≈300 bursts from 1878 to 2025 to train the model; the train-to-test ratio was 75%–25%. Based on cross validation, we find that the model achieves MAEs of approximately 7 SSN for amplitude and 1.8 months for duration. These results demonstrate that the RF layer significantly reduces forecast bias that would otherwise be introduced by purely autoregressive filters. In addition, the machine learning component generates uncertainty estimates, which we incorporate directly into our probabilistic space-weather forecasting products.

Lastly, we apply a multiplicative damping factor to both the predictive strategies and the 12 month exponential half-life imposed on postmaximum bursts. These conditions are consistent with empirical cycle 24 and 25 decline rates (e.g., L. A. Upton & D. H. Hathaway 2023). While ad hoc in appearance, these scalings mimic the declining free energy in the convection zone after the global maximum and prevent overforecasting late cycle activity, a common failure mode of linear SARIMA extrapolations. Together, these tailored models narrow the intermediate-range forecasting gap that has long limited operational space-weather planning. Key forecast highlights are as follows.

1. *Northern Hemisphere*. A hindcast burst runs from 2024 October to 2025 June and peaks in 2025 March at ≈70–80 SSN; the next forecast burst is expected between 2025 October and 2026 March, peaking in 2025 December at ≈55–65 SSN.
2. *Southern Hemisphere*. The dominant burst spans late 2024 through mid-2025, cresting in 2025 March–April at ≈130–140 SSN; a tertiary burst is projected for 2025 December–2026 April, peaking near 2026 January at ≈120 SSN.
3. *Total disk*. The combined hemispheric signal is forecast to reach ≈205–210 SSN in 2025 March–April, then plateau at ≈170–175 SSN through late 2025 before entering a gradual decline.

## Acknowledgments

This work is supported by the NSF grant 1936336 and 2401175 and NASA grant NNH23ZDA001N-BPSF. In addition, the work is supported by the NSF National Center for Atmospheric Research, which is a major facility sponsored by the National Science Foundation under cooperative agreement 1852977. We also acknowledge support from several NASA grants, namely M. D. acknowledges NASA-HSR award 80NSSC21K1676, Stanford COFFIES Phase II NASA-DRIVE Center subaward 80NSSC22M0162, and NASA-HSR subaward 80NSSC21K1678 from JHU/APL. This research was made possible by the open-source projects `matplotlib`, `numpy`, and `scipy`.

## Author Contributions

All authors contributed equally to this work.

## Appendix A
## Gaussian Fitting Routines

We analyze monthly hemispheric SSNs using Gaussian smoothing with window FWHM $W \in \{2, 3, 4, 6, 8\}$ months, applied directly to the monthly series without detrending. Peaks are detected on each smoothed curve with a minimum separation of 2 months and an SSN peak threshold of six;

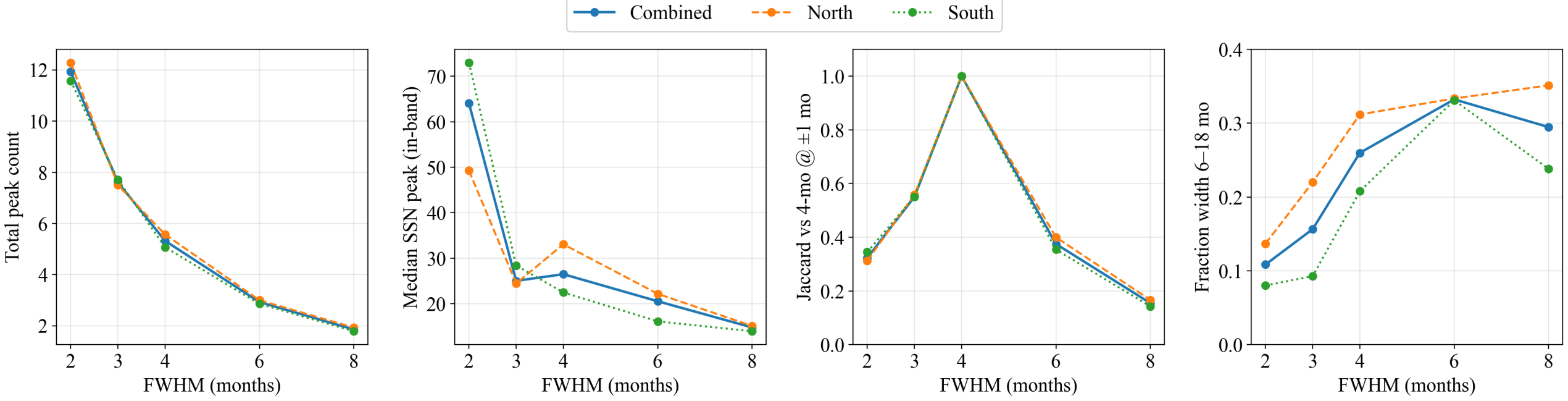

**Figure A1.** Zoomed-in burst-envelope forecasts from 2024 January to 2026 April. Top: the Northern Hemisphere SSN with inferred RF burst properties. Middle: the Southern Hemisphere counterpart. Bottom: total (Northern Hemisphere + Southern Hemisphere) SSN, highlighting the same burst timings and amplitudes.

widths are measured at half prominence and, because the sampling is monthly, the widths in samples equal widths in months. The Gaussian scale used by the filter is related to the chosen window by

$$\sigma = \frac{W}{2\sqrt{2\ln 2}}, \tag{A1}$$

and a peak is classified as "burst-like" when its width lies between the 6 and 18 month band.

Figure A1 summarizes the window sensitivity across cycles 12–25 for the combined series (solid) and the Northern and Southern Hemispheres (dashed–dotted). Panel (a) shows that the total peak count drops sharply as $W$ increases, indicating that 2–3 month windows oversegment the series into many narrow peaks. Panel (b) reports the median SSN peak for events within 6–18 months; values are robust at $W = 4$ months and decline for longer windows as nearby features are merged and broadened. Panel (c) gives the Jaccard overlap between peak times for each window and the 4 month reference (tolerance $\pm$ 1 month); the overlap is maximal at 4 months and decreases at 6–8 months, evidencing timing drift/merging with heavier smoothing. Panel (d) shows the fraction of peaks with widths of 6–18 months; this fraction rises from 2–3 to ~6–8 months as month-scale undulations are suppressed. Together, these diagnostics support the use of a 4 month Gaussian window as a balanced choice: it removes month-scale variability while preserving the morphology, amplitude, and timing of burst-scale (6–18 months) features.

## Appendix B
## Uncertainty Quantification and Model Evaluation

The hybrid nature of our forecasting framework means that the total uncertainty in our final predictions (Figure 6) is a composite of several distinct sources. Our model evaluation relies on a suite of statistical metrics: including the AIC for model selection, and MAE and RMSE for accuracy assessment to ensure robustness. Here, we detail the components of uncertainty to provide a transparent framework for interpreting our forecast's reliability.

The foundational layer of uncertainty originates from the SARIMA baseline model. We quantify this using a 95% CI, visualized as the shaded region in Figure 6. For a forecast $k$ steps into the future, the interval is calculated based on the forecast value ($\hat{y}_{t+k}$) and the standard error of the forecast variance ($\sigma_k^2$), which accounts for uncertainty in both the model parameters and the inherent stochasticity of the time series. The 95% CI is given by

$$\mathrm{CI}(k) = \hat{y}_{t+k} \pm 1.96 \cdot \sigma_k, \tag{B1}$$

where 1.96 is the critical value obtained from the standard normal distribution for a 95% confidence level. This interval represents the range within which the true SSN is expected to fall with 95% probability, assuming the model is correctly specified.

The second layer of uncertainty, which pertains to the individual bursts predicted by the RF model, is represented graphically using a pragmatic, data-driven proxy. Unlike the statistically derived CI for the SARIMA baseline, the uncertainty for the RF predictions of burst amplitude and duration is visualized based on heuristics defined in our analysis scripts. The vertical error bars shown for each forecast burst in Figure 6 are computed as a fixed fraction (15%) of the predicted burst amplitude. This value is defined within our visualization script to serve as a proxy for the 1$\sigma$ uncertainty, reflecting a reasonable assumption that larger bursts are subject to greater variability. The horizontal shaded regions, which denote the burst's effective duration, are visualized to span 50% of the predicted rise and decay periods. While not a formal prediction interval from the RF model itself, this visualization provides crucial physical context by showing how the predicted burst parameters compare to the range of historically observed events.

Finally, a third, structural uncertainty exists in our assumption that solar bursts are well modeled by asymmetric Gaussian profiles. A formal propagation of these uncertainty layers is nontrivial, as the RF model's inputs are derived from the SARIMA forecast, making the baseline and burst uncertainties interdependent. A rigorous propagation would require computationally intensive methods, such as Monte Carlo simulations, which represent a valuable direction for future work.

## ORCID iDs

Juie Shetye https://orcid.org/0000-0002-4188-7010
Mausumi Dikpati https://orcid.org/0000-0002-2227-0488

## References

Box, G., & Jenkins, G. 1976, Time Series Analysis: Forecasting and Control (San Francisco, CA: Holden-Day)
Breiman, L. 2001, MachL, 45, 5
Clette, F., Svalgaard, L., Vaquero, J. M., & Cliver, E. W. 2014, SSRv, 186, 35
Dash, S., DeRosa, M. L., Dikpati, M., et al. 2024, ApJ, 975, 288

Dikpati, M., Cally, P. S., McIntosh, S. W., & Heifetz, E. 2017, NatSR, 7, 14750
Dikpati, M., Belucz, B., Gilman, P. A., & McIntosh, S. W. 2018a, ApJ, 862, 159
Dikpati, M., McIntosh, S. W., Bothun, G., et al. 2018b, ApJ, 853, 144
Dikpati, M., & McIntosh, S. W. 2020, SpWea, 18, e2018SW002109
Dikpati, M., McIntosh, S. W., Chatterjee, S., et al. 2021, ApJ, 910, 91
Dikpati, M., Gilman, P. A., Guerrero, G. A., et al. 2022, ApJ, 931, 117
Dikpati, M., Raphaldini, B., McIntosh, S. W., et al. 2024, PNAS, 121, e2415157121
Dikpati, M., Korsós, M. B., Norton, A. A., et al. 2025, ApJ, 988, 108
Gilman, P. A. 2000, ApJL, 544, L79
Gurgenashvili, E., Zaqarashvili, T. V., Kukhianidze, V., et al. 2016, ApJ, 825, 55
Gurgenashvili, E., Zaqarashvili, T. V., Kukhianidze, V., et al. 2017, ApJ, 845, 137
Gurgenashvili, E., Zaqarashvili, T. V., Kukhianidze, V., et al. 2022, A&A, 660, A33
Hathaway, D. H. 2015, LRSP, 12, 4
Hintze, J. L., & Nelson, R. D. 1998, The American Statistician, 52, 181
Hyndman, R. J., & Athanasopoulos, G. 2018, Forecasting: Principles and Practice (2nd ed.; Melbourne: OTexts) https://OTexts.com/fpp2/
Inceoglu, F., Simoniello, R., Arlt, R., & Rempel, M. 2019, A&A, 625, A117
Lobzin, A., Cairns, I. H., & Robinson, P. A. 2012, ApJL, 754, L28
McIntosh, S. W., & Leamon, R. J. 2014, ApJL, 796, L19
McIntosh, S. W., Leamon, R. J., Krista, L. D., et al. 2015, NatCo, 6, 6491
McIntosh, S. W., Cramer, W. J., Marcano, P., & Leamon, R. J. 2017, NatAs, 1, 86
Norton, A. A., & Gallagher, J. C. 2010, SoPh, 261, 193
Paul, K. S., Moses, M., Haralambous, H., & Oikonomou, C. 2025, RemS, 17, 859
Raphaldini, B., Dikpati, M., Norton, A. A., et al. 2023, ApJ, 958, 175
Rieger, E., Share, G. H., Forrest, D. J., Kanbach, D., & Chupp, E. L. 1984, Natur, 312, 623
Schrijver, C. J., Kauristie, K., Aylward, A. D., et al. 2015, AdSpR, 55, 2745
Temmer, M., Rybák, J., Bendík, P., et al. 2006, A&A, 447, 735
Upton, L. A., & Hathaway, D. H. 2023, JGRA, 128, e2023JA031681
Vecchio, A., & Carbone, V. 2009, A&A, 502, 981
Xu, Q., Jain, R., & Xing, W. 2024, SoPh, 299, 25
Zaqarashvili, T. V., Oliver, R., Ballester, J. L., & Shergelashvili, B. M. 2007, A&A, 470, 815
Zaqarashvili, T. V., Carbonell, M., Oliver, R., & Ballester, J. L. 2010, ApJ, 709, 749
Zhu, H., Zhu, W., & He, M. 2022, SoPh, 297, 157